\documentstyle[epsfig]{mn}

\def\i{\item}
\def\bi{\bibitem{}}

\def\ni{\noindent}
\def\beb{\begin{thebibliography}{999}}
\def\eeb{\end{thebibliography}}
\def\bei{\begin{itemize}}
\def\eei{\end{itemize}}
\def\bef{\begin{figure}}
\def\eef{\end{figure}}
\def\ben{\begin{enumerate}}
\def\een{\end{enumerate}}
\def\beq{\begin{equation}}
\def\eeq{\end{equation}}
\def\ber{\begin{eqnarray}}
\def\eer{\end{eqnarray}}
\def\edo{\end{document}}

\newcommand{\dmdt}{{\mbox{{\rm M}$_{\odot}$}} {\rm yr}$^{-1}$}

\newcommand{\gcc}{{\rm g} \, {\rm cm}^{-3}}

\newcommand{\msun}{\mbox{{\rm M}$_{\odot}$}}
\newcommand{\mdot}{\mbox{$\dot{M}$}}

\begin{document}
\title[neutron star field evolution]
{Magnetic Field Evolution of Accreting Neutron Stars - IV\\
{Effect of the Curvature of Space-Time}} 
\author[Konar]
{Sushan Konar$^{\dag}$ \\ 
Inter-University Centre for Astronomy \& Astrophysics, Pune 411007, India \\ 
$^{\dag}$ e-mail : sushan@iucaa.ernet.in \\ }
\date{2nd May, 2000}
\maketitle

\begin{abstract}
The evolution of the magnetic field in an accreting neutron star is investigated using a fully general relativistic 
treatment and assuming that initially the currents supporting the field are completely confined to the crust. We 
find that the field decay slows down due to the inclusion of the curvature of space-time but the final results do not 
differ significantly from those obtained assuming a flat space-time. We also find that such modifications are small 
compared to the uncertainties introduced by a lack of precise knowledge of the neutron star micro-physics.
\end{abstract}


\section{introduction}
\label{sintro}

\ni The generation and evolution of the magnetic field in neutron stars continue to evoke much interest, particularly 
because there is as yet no consensus on the question of field generation. The field could be a fossil remnant from the 
progenitor star in the form of Abrikosov fluxoids embedded in the proton superconductor of the core (Baym, 
Pethick \& Pines 1969, Ruderman 1972). Or it could be generated after the formation of a neutron star in which case 
the currents would be entirely confined to the solid crust (Blandford, Applegate \& Hernquist 1983, Urpin, Levshakov 
\& Yakovlev 1986). Alternatively, a seed magnetic field could be amplified due to the entropy-driven convection in a 
young neutron star (Thompson \& Duncan 1993). Observations and statistical analyses of existing pulsar data, however, 
indicate that significant decay of magnetic field is achieved only if the neutron star is a member of an interacting 
binary (see Bhattacharya 1995, 1996, van den Heuvel 1995, Verbunt \& van den Heuvel 1995 and references therein) 
irrespective of the generation mechanism. \\

\ni The processes responsible for the field evolution, in neutron stars which are members of binaries, are - a) expulsion
of the magnetic flux from the superconducting core during the phase of propeller spin-down, b) rapid ohmic decay of the 
crustal field in an accretion-heated crust and c) screening of the field by accreted matter. Amongst these, the diamagnetic 
screening of the field by accreted matter depends on the detailed geometry of the material flow and whether it can have
a long-term effect is still not very clear (Konar 1997, Choudhuri \& Konar 2002). The rest of the models invoke ohmic decay 
of the current loops, residing in the crust, for a permanent decrease in the field strength (see Konar \& Bhattacharya 2001 
for a brief review). In either case, the effect of accretion is two-fold. The heating reduces the electrical conductivity,
and consequently the ohmic decay time-scale, inducing a faster decay. At the same time the material movement, caused by the 
deposition of matter on top of the crust, pushes the original current carrying layers into deeper and denser regions where 
the higher conductivity slows the decay down. \\

\ni Though the problem of the evolution of the magnetic field in neutron stars have been investigated by many authors a 
fully general relativistic treatment has not been attempted by many. Recently, Sengupta (1997, 1998) addressed this issue 
for isolated pulsars and found that the modification introduced by a general relativistic treatment is significantly large. 
However, our results are at variance with this claim (Ramachandra \& Vishweshwara 2000). Indeed this is supported by the 
recent work of Page, Geppert \& Zannias (2000) and Geppert, Page \& Zannias (2000). These authors have investigated coupled 
spin, magnetic field and thermal evolution of isolated neutron stars in a general relativistic framework. And they find that 
the change in the final field strength is quite small. \\

\ni In a series of papers (Konar \& Bhattacharya 1997, 1999a, 1999b) we have discussed the accretion induced field 
evolution in neutron stars. The formalism adopted in these papers implicitly assumes a flat space-time. In this work we 
report the results of our investigation of the field evolution in an accreting neutron star adopting a general 
relativistic framework. Our aim is to calculate the amount of modification introduced by assuming a curved space-time 
vis-a-vis the uncertainties already existing in the field evolution calculation assuming a flat space-time. It is well 
known that such uncertainties arise out of a lack of knowledge regarding the values of various crustal parameters (like 
the density at which the currents are concentrated or the impurity content of the crust) and the observations are still 
not adequate to define those with better precision. \\

\ni In this work we assume that the currents supporting the field are initially completely confined to the crust. This 
assumption does not affect the general qualitative conclusions obtained from our present calculation. Because,it has 
already been shown (Konar \& Bhattacharya 1999b) that an initial configuration supported by the magnetic fluxoids in 
the proton superconductor behaves in a similar fashion once it is converted to crustal currents by spin-down driven 
flux expulsion. \\

\ni We also assume the mass accretion to be spherical. Of course, except for very large values of mass transfer the accretion 
would be channeled through the polar cap. The detailed nature of this material flow and its effect on the magnetic field has 
recently been investigated (Choudhuri \& Konar 2002, Konar \& Choudhuri 2002). But such flows are confined only to the layers 
very close to the surface. But the currents would normally be concentrated in the deeper curtal layers for the magnetic field
to be long lived. And in these layers the material movement would be purely radial due to the overall compression of the neutron 
star, as a result of an increase in the mass. Therefore, an assumption of a spherical accretion would not introduce any 
significant error in our present calculations. \\

\ni The organisation of the paper is as follows. In section \ref{sinduct} we derive the covariant form of the induction
equation that governs the temporal evolution of the magnetic field. In section \ref{scrust} we discuss the detailed form
of the induction equation, both for a flat and a curved space-time, in the crust of a neutron star. Section \ref{sns} 
provides some details of the microphysics of the neutron star required for our investigation. In section \ref{sresults} 
we present the results of our calculation and finally conclude in section \ref{sconcl}.

\section{the induction equation}
\label{sinduct}

The induction equation in a flat space-time is given by (Jackson 1975):
\beq
\frac{\partial \vec {\cal B}}{\partial t} = \vec \nabla \times (\vec V \times \vec {\cal B}) 
- \frac{c^2}{4 \pi} \vec \nabla \times (\frac{1}{\sigma} \times \vec \nabla \times \vec {\cal B}) \,,
\label{e-ind}
\eeq
where $\vec V$ is the velocity of material movement and $\sigma$ is the electrical conductivity of the medium.  
To obtain the covariant form of the induction equation we make use of the covariant form of the Maxwell's 
equations (Weinberg 1972):
\ber
\frac{1}{\sqrt{-g}}\partial_{\mu} ( \sqrt{-g} F^{\mu \nu}) &=& - \frac{4 \pi}{c} \, J^{\nu}\,, \\
\partial_{\mu} F^{\nu \lambda} + \partial_{\nu} F^{\lambda \mu} + \partial_{\lambda} F^{\mu \nu} &=& 0 \,;
\eer
with 
\ber
F_{\mu \nu} &=& \partial_{\mu} A_{\nu} - \partial_{\nu} A_{\mu} \,,\nonumber \\
J^{\mu} &=& (c\rho, {\bf J}) \,, \nonumber 
\eer
where $A_{\mu}$ is the vector potential, $\rho$ is the charge density and ${\bf J}$ is the spatial part of the
current density. Using the generalised Ohm's law given by (Weinberg 1972):
\beq
J^{\mu} = \sigma g^{\mu \nu} F_{\nu \lambda} u^{\lambda} \,,
\eeq
where $u^{\lambda}$ is the covariant velocity, from the first Maxwell's equation we obtain:
\beq
F_{\alpha 0} 
= - \frac{c}{4 \pi \sigma} \frac{1}{\sqrt{-g}} g_{\alpha \nu} \partial_{\mu} ( \sqrt{-g} F^{\mu \nu}) 
- F_{\alpha k} \frac{u^k}{u^0} \,.
\eeq
With the covariant velocity given by $ u^{\mu} = (\frac{dx^0}{ds}, \frac{1}{c} u^0 {\bf V})$, where ${\bf V}$ is 
the velocity in the locally inertial frame, we obtain from the second Maxwell's equation:
\ber
\partial_t F_{i j} &=& \partial_{i} (F_{j k} v^k) - \partial_{j} (F_{i k} v^k) \nonumber \\
&& + \frac{c^2}{4 \pi} \partial_{i} \left(\frac{1}{\sigma} \frac{1}{\sqrt{-g}} \frac{g_{i \nu}}{u^0} 
\partial_{l} (\sqrt{-g} F^{l \nu})\right) \nonumber \\
&& - \frac{c^2}{4 \pi} \partial_{j} \left(\frac{1}{\sigma} \frac{1}{\sqrt{-g}} \frac{g_{j \nu}}{u^0} 
\partial_{l} ( \sqrt{-g} F^{l \nu})\right) \,,
\label{e-coind}
\eer
where $x^{\mu} = (ct, r, \theta, \phi)$.
It should be noted that we have neglected the displacement current here and $i,j,k,l = 1,2,3$ since we are only
interested in the time evolution of the magnetic field. For a given $F_{\mu \nu}$ an observer with four velocity
$u^{\mu}$ measures the electric and the magnetic fields $(E,B)$ as given by:
\beq
E_{\mu} = F_{\mu \nu} u^{\nu} 
\; \; \mbox{and} \; \;B_{\mu} = - \frac{1}{2} {\epsilon_{\mu \nu}}^{\gamma \delta} F_{\gamma \delta} u^{\nu},
\label{e-field}
\eeq
where $\epsilon_{\mu \nu \gamma \delta}$ is the four-dimensional Levi-Civita tensor (Wald 1984). Therefore 
eq.(\ref{e-coind}) gives the covariant form of the induction equation which reduces to eq.(\ref{e-ind}) in the 
limit of a flat metric. We must also mention here that in this work we assume that the magnetic field is weak 
enough for the electrical conductivity to be isotropic.

\section{evolution in the crust of a neutron star}
\label{scrust}

\ni In this work we shall follow the formalism developed by Konar \& Bhattacharya (1997 - paper I hereafter) for the
evolution of a crustal field assuming a flat space-time. And shall compare the results with those obtained in paper I
to obtain the modifications introduced by assuming a curved space-time. \\

\ni We assume the star to be accreting at an uniform rate, without any loss of generality, since the purpose of this 
work is to find the modification introduced by the curvature of space-time. Therefore we shall, for the moment, ignore 
the realistic situation of a neutron star accreting at different rates as its binary companion goes through different
stages of evolution. Assuming the mass flow to be spherically symmetric in the crustal layers of interest, the velocity 
of the material movement is given by (see paper I for details):
\beq
v(r) = -\frac{\mdot}{4\pi^2\rho(r)}\,,
\eeq
where $\mdot$ is the rate of mass accretion and $\rho(r)$ is the density as a function of radius $r$.\\

\ni The mass in the crust of the neutron star is quite small compared to the total mass of the star. For example for a 
intermediate equation of state (Wiringa, Fiks \& Fabrocini 1988 matched to Negele \& Vautherin 1973 and Baym, Pethick \& 
Sutherland 1971) the mass of the crust of a 1.4 \msun neutron star would be 0.044 \msun, which is $\sim 3\%$ of the 
total mass. Therefore, in the crust of a non-rotating neutron star we can use the exterior Schwarzschild metric given by
(Weinberg 1972):
\beq
g_{\mu \nu} = \left( -(1 - \frac{2m}{r}), (1 - \frac{2m}{r})^{-1}, r^2, r^2 \sin^2 \theta \right) 
\label{e-metric}
\eeq
without making any significant error. Here, $m = \frac{G M}{c^2}$, $M$ being the mass of the star. \\

\ni It should be noted here that the crustal mass of a neutron star remains effectively constant for the accreted masses 
of the order of 0.1~\msun (in fact, with a slight decrease) as the total mass increases. Therefore accretion implies 
assimilation of the original crust into the superconducting core. We assume that when the original current carrying 
regions undergo such assimilation the superconducting transition happnes over a time-scale much smaller than the 
flux-expulsion time-scale and hence the flux remains {\em frozen} thereafter stopping any further decay.  Since we 
terminate our calculation long before the crustal mass changes significantly (as the field `levels off' due to 
{\em flux freezing} much before that) the use of exterior Schwarzschild metric remains valid even in the crust of an 
accreting neutron star. \\

\subsection{Flat Space-Time}
\label{sscst}

In a flat space-time :
\beq
\vec {\cal B} = \nabla \times \vec {\cal A}\,,  
\eeq
where we assume ${\vec {\cal A}} = (0, 0, {\cal A}_{\phi})$. This choice of the vector potential ensures a poloidal
geometry for $\vec {\cal B}$. In particular, we take,
\beq
{\cal A}_{\phi} = \frac{g(r,t) \sin \theta}{r}\,, 
\eeq
where $g(r,t)$ is the Stokes' function. In the lowest order of multipole, the dipolar form of the magnetic field is 
given by:
\ber
{\cal B}_r &=& \frac{2 \cos \theta g(r,t)}{r^2} \,,\\
{\cal B}_{\theta} &=& - \frac{\sin \theta}{r}\frac{\partial g(r,t)}{\partial r} \,;
\eer
where, the quantities ${\cal A, B}$ correspond to those in a flat space-time. The induction equation then takes the form:
\beq
\partial_t g(r,t) = \frac{c^2}{4 \pi \sigma} \left\{\partial^2_r g(r,t) - \frac{2 g(r,t)}{r^2} \right\} + \partial_r g(r,t) v(r).
\label{e-dgdtf}
\eeq
At the pole the magnitude of the magnetic field is proportional to $\frac{2 g(R,t)}{R^2}$. Therefore, to find the 
time-evolution of the strength of the surface magnetic field at the pole we solve eq.(\ref{e-dgdtf}) subject to the 
following boundary conditions (see e.g., Geppert \& Urpin 1994):
\ber
\frac{\partial g(r,t)}{\partial r}|_{r=R} + \frac {g(R,t)}{R} &=& 0,\\
\label{e-bc1}
g(r_{\rm co},t) &=& 0
\label{e-bc2}
\eer
where $R$ is radius of the star and $r_{\rm co}$ is that radius to which the original boundary between the core and the 
crust is pushed to, due to accretion, at any point of time. The first condition matches the interior field to an external 
dipole configuration. The second condition indicates that as accretion proceeds along with the crustal material the 
frozen-in flux moves inside the core. For the details regarding the particular form of $g(r,0)$ and the current density
corresponding to it, please see paper I.

\subsection{Curved Space-Time}
\label{sscst}
\ni In a Schwarzschild metric $F_{\mu \nu}$ is related to the locally measured magnetic magnetic field ${\cal B}$ by the
following relations (Wasserman \& Shapiro 1983):
\ber
{\cal B}_{\theta} &=& \frac{(1 - 2m/r)^{1/2}}{r \sin \theta} \, F_{\phi r}, \\
{\cal B}_r &=& - \frac{1}{r^2 \sin \theta} \, F_{\phi \theta}.
\eer
Then the choice of, $A_{\phi} = g(r,t) \sin^2 \theta$, ensures that $B$ and ${\cal B}$ match in a locally flat space-time
where $B$ is obtained from eq.(\ref{e-field}). With this choice, we have (Wasserman \& Shapiro 1983):
\ber
B_r &=& \frac{2 \cos \theta}{r^2} \, g(r,t) \,,\\
B_{\theta} &=& - \left(1 - \frac{2m}{r}\right)^{1/2} \,\frac{\sin \theta}{r} \, \partial_r g(r,t) \,.
\eer
Therefore, in a curved space-time the induction equation reduces to:
\ber
\partial_t g(r,t) &=& \frac{c^2}{4 \pi \sigma} \left(1 - \frac{2m}{r}\right)^{1/2} 
\Big[\left(1 - \frac{2m}{r}\right) \partial^2_r g(r,t) \nonumber \\
&& - \frac{2 g(r,t)}{r^2} \Big] \nonumber \\
&& + \Big[ \frac{c^2}{4 \pi \sigma} \, \left(1 - \frac{2m}{r}\right)^{1/2} \, \frac{2m}{r^2} 
+ v(r) \Big] \partial_r g(r,t). \nonumber \\
&&
\label{e-dgdtc}
\eer
Once again the strength of the field at the pole is simply proportional to $g(R,t)$. But now the boundary conditions
get modified to the following form (see Zannias, Geppert \& Page 2000):
\ber
\partial_r g(r,t)|_{r=R} + {\cal G} \frac {g(R,t)}{R} = 0,\\
g(r_{\rm co},t) = 0\,;
\eer
with,
\beq
{\cal G} = y \frac{2y \ln (1 - y^{-1}) + \frac{2y - 1}{y-1}}{y^2 \ln (1 - y^{-1}) + y + 1/2}, \, \mbox{$y=R/2m$}.
\eeq

\section{neutron star micro-physics}
\label{sns}

\subsection{Equation of State}
\label{sseos}

\ni In order to investigate the effect of the curvature of space-time on the evolution a purely crustal field,
we consider three equations of state - soft, intermediate and stiff. The soft eos is taken from Pandharipande
(1971) (P71 hereafter), the intermediate from Wiringa, Fiks \& Fabrocini (1988) (WFF hereafter) and the
stiff eos from the work by Walecka (1974) (W74 hereafter) respectively. All these eos are matched to Negele
\& Vautherin (1973) and Baym, Pethick \& Sutherland for the low-density crustal region. In table (\ref{t-radius}) 
we have listed the radii of stars of different masses obtained by using different equations of state. 

\begin{table}
\begin{center}{
\begin{tabular}{|l|l|r|r|} \hline
M/\msun & P71 & WFF 2 & W74 \\ \cline{1-4}
1.4 & 7.45 & 11.08 & 12.21 \\ \cline{1-4}
1.6 & & 11.03 & 12.18 \\ \cline{1-4}
1.8 & & 10.91 & 12.10 \\ \cline{1-4}
2.0 & & & 11.91 \\ \cline{1-4}
\end{tabular}}
\end{center}
\caption[]{Radii of neutron stars of different masses for different equations of state, in Km. The equations
of state P71, WFF and W74 have been discussed in the text.}
\label{t-radius}
\end{table}

\subsection{Crustal Physics}
\label{sscp}

\ni As has been discussed in paper I for the uniform crustal temperature of a neutron star, accreting at a rate
of \mdot, we use a fitting formula to the results obtained by Zdunik et al. (1992) given by:
\beq
\log T = 0.397 \log \mdot + 12.35\,.           
\label{e-temp}
\eeq
Extrapolation of this fit to higher rates of accretion gives un-physically high temperatures. It has been shown by
Brown (2000) that the maximum temperature obtained in an accretion heated crust is $\sim 10^{8.5}$~K. Therefore we 
use this value if the temperature obtained through eq.(\ref{e-temp}) exceeds it. It should be noted here that the 
above results were obtained for a neutron of mass 1.4~\msun with an intermediate equation of state. Since, such 
calculations do not exist for different masses or different equations of state, use of eq.(\ref{e-temp}) for all 
situations introduce some error in our calculation. But since the purpose of this work is to obtain a measure of 
the modification in the final field strength by using a general relativistic framework we shall, for the moment, 
use eq.(\ref{e-temp}) as an approximate indicator for the crustal temperature.\\

\ni In the solid crust, the electrical conductivity has contributions from both the phonon and the impurity processes and is
obtained as,
\beq
\frac{1}{\sigma} = \frac{1}{\sigma_{\rm ph}} + \frac{1}{\sigma_{\rm imp}}\,,
\label{e-sigma}
\eeq
where $\sigma_{\rm ph}$ is the phonon scattering conductivity dependent upon the density and the temperature 
(Itoh et al. 1984) and $\sigma_{\rm imp}$ is the impurity conductivity dependent upon the density and the impurity 
parameter $Q$ (Yakovlev \& Urpin 1980). Whereas, $\sigma_{\rm ph}$ and $\sigma_{\rm imp}$ are given by,
\ber
\sigma_{\rm ph}  &=& 1.24 \times 10^{20} \frac{x^{4}}{u T_{8}}
               \frac{(u^{2} + 0.0174)^{1/2}}{(1 + 1.018 x^{2}) I_{\sigma}} \, s^{-1}, \\
\sigma_{\rm imp} &=& 8.53 \times 10^{21} x  Z/Q \, s^{-1}
\eer
with,
\ber
u &=& \frac{2 \pi}{9} (log \rho - 3) \nonumber \\
T_{8} &=& \mbox{temperature in units of $10^8$ K} \\
\rho_{6} &=& \mbox{density in units of $10^6~\gcc$} \\
I_{\sigma} &=& \mbox{a function of $\rho, Z, A$}.           
\eer
The temperature or density of the cross-over from phonon dominated to impurity dominated process depends on the impurity 
strength $Q$, given by,
\beq
Q = \frac{1}{n} \sum_{i}{{n_{i}}(Z - Z_{i})^2} 
\eeq
where $n$ is the total ion density, $n_i$ is the density of impurity species $i$ with charge $Z_i$, and $Z$ is the ionic 
charge in the pure lattice (Yakovlev \& Urpin 1980). \\

\bef
\begin{center}{\mbox{\epsfig{file=fig01.ps,width=225pt}}}\end{center}
\caption[]{The evolution of the surface magnetic field in an accreting neutron star of mass 1.4~\msun  corresponding to
an intermediate equation of state given by WFF, assuming a flat space-time. Legends next to the curves - {\bf 1} and {\bf 2} 
correspond to $\mdot = 10^{-12}, 10^{-10}$~\msun/yr respectively, {\bf A} and {\bf B} correspond to densities of 
$10^{14}, 10^{13}~\gcc$ respectively (at which the initial current configurations are centred) and {\bf a} and {\bf b} 
correspond to impurity contents of 0 and 0.01 respectively.}
\label{fig01}
\eef

\ni It should be mentioned here that in an accretion heated crust the effect of the impurities is negligible except when 
the rate of mass accretion is small implying a lower crustal temperature. This is true even when the crustal currents are 
concentrated at higher densities. To demonstrate this, we present some calculations for different values of $Q$ in 
fig.(\ref{fig01}). In this plot the evolution of the surface field with time has been shown for two different values of 
the impurity concentration. The calculations pertain to two different rates of accretion and for currents concentrated at 
two different densities. It is evident from the curves that the impurity content is important only for lower rates of mass 
accretion. It should be noted that in fig.(\ref{fig01}) the curves {\bf 2Aa, 2Ab} and {\bf 2Ba, 2Bb} are actually 
pair-wise indistinguishable signifying that the effect of the impurity content is completely insignificant for 
somewhat larger rates of accretion. Since, in the present work we concentrate only on demonstrating the effect of space-time 
curvature on the evolution of the magnetic field, $Q=0$ is assumed for the rest of our calculations.

\section{results and discussions}
\label{sresults}

\subsection{Final Surface Fields}
\label{ssfinal}

\ni We summarise the results of our investigation in the following plots. In fig.(\ref{fig02}) we plot the
evolution of the surface magnetic field for various densities of current concentration and for a fixed rate of
mass accretion. It is clearly seen that the difference in final field values between that obtained using a flat
and a curved space-time is $\sim$ 1/4 orders of magnitude whereas that obtained by using two different densities 
for the initial current concentration could be as large as an order of magnitude. Similarly in fig.(\ref{fig03}) 
the difference in the final field values obtained using two different rates of accretion is much larger than any 
modification introduced by a general relativistic framework. Since the values of the current concentration density 
or the rate of accretion used by us are well within the limits of observational uncertainty the inclusion/exclusion 
of a general relativistic framework does not really affect the calculation of the evolution of crustal magnetic field. \\

\bef
\begin{center}{\mbox{\epsfig{file=fig02.ps,width=225pt}}}\end{center}
\caption[]{The evolution of the surface magnetic field in an accreting neutron star of mass 1.4~\msun  corresponding to
an intermediate equation of state given by WFF. Curves 1 to 3 correspond to densities of $10^{13}, 10^{12}, 10^{11}~\gcc$ 
respectively, at which the initial current configurations are centred, assuming a flat space-time. Whereas, curves 1a to 3a 
correspond to curved space-time. All curves correspond to $\mdot = 10^{-10}$~\msun/yr and $Q$ = 0.}
\label{fig02}
\eef

\bef
\begin{center}{\mbox{\epsfig{file=fig03.ps,width=225pt}}}\end{center}
\caption[]{The evolution of the surface magnetic field in an accreting neutron star of mass 1.4~\msun  corresponding to
an intermediate equation of state given by WFF. Curves 1 to 3 correspond to $\mdot = 10^{-12}, 10^{-11}, 10^{-10}$~\msun/yr 
respectively, assuming a flat space-time. Curves 1a to 3a correspond to curved space-time. All curves correspond to 
$Q$ = 0 and a density of $10^{11}~\gcc$ at which the current distribution is initially concentrated.}
\label{fig03}
\eef

\bef
\begin{center}{\mbox{\epsfig{file=fig04.ps,width=225pt}}}\end{center}
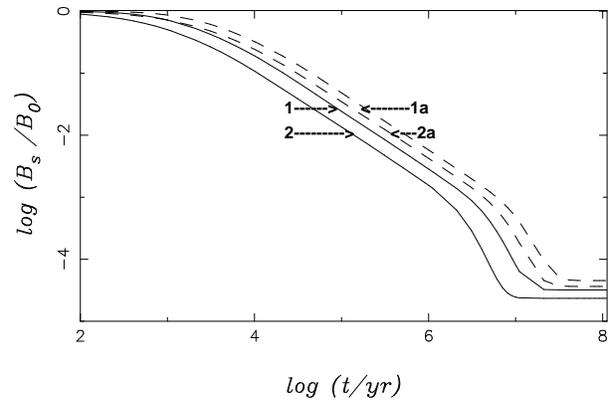
\caption[]{The evolution of the surface magnetic fields in neutron stars accreting at a rate $\mdot = 10^{-10}$~\msun/yr,
corresponding to an intermediate equation of state given by WFF. Curves 1 and 2 correspond to the total masses of the
neutron star being equal to 1.4 and 1.8~\msun respectively, assuming a flat space-time. Curves 1a and 2a correspond to 
curved space-time. All curves correspond to $Q$ = 0 and a density of $10^{11}~\gcc$ at which the current distribution 
is initially concentrated.}
\label{fig04}
\eef

\bef
\begin{center}{\mbox{\epsfig{file=fig05.ps,width=225pt}}}\end{center}
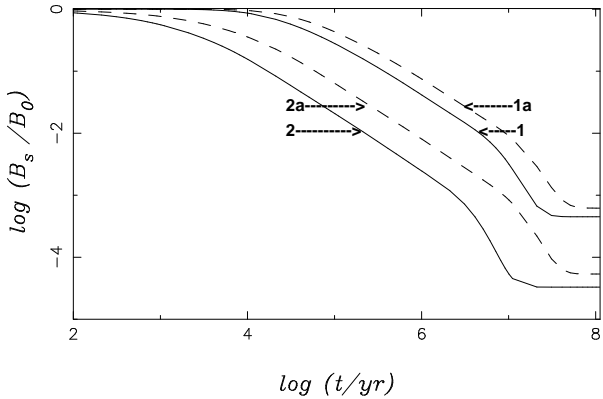
\caption[]{The evolution of the surface magnetic fields in neutron stars, of mass 1.4~\msun,  accreting at a rate 
$\mdot = 10^{-10}$~\msun/yr. Curves 1 and 2 correspond to W74 and P71 equations of state respectively, assuming a 
flat space-time. Curves 1a and 2a correspond to curved space-time. All curves correspond to $Q$ = 0 and a density 
of $10^{11}~\gcc$ at which the current distribution is initially concentrated.}
\label{fig05}
\eef

\ni In fig.s (\ref{fig04}) and (\ref{fig05}) we compare the effect of curved space-time for neutron stars of different
masses and for different equations of state. The actual amount of field decay is more for a compact (a stiffer equation 
of state or a larger mass) star owing to a thinner crust and hence a smaller diffusive length-scale. But the modification
due to general relativistic effects is larger for compact stars. That effect is discernible in fig.(\ref{fig05}). But the 
effect of a mass difference is rather insignificant as is seen in fig.(\ref{fig04}). It should be noted here that the
crustal models used in our calculation (described in section~\ref{sseos}) are somewhat older. The newer models predict
lower crustal mass (Lorenz, Ravenhall \& Pethick 1993) implying a higher surface red-shift for a given total mass of
the star. It is evident from fig.(\ref{fig05}) that such an effect is again small compared to other factors mentioned
above.

\subsection{Effect of the Curvature of Space-Time}
\label{sseffect}

\ni To understand the effect of the curvature of space-time on the evolution of a purely crustal field we
first define the characteristic diffusive and advective time-scales for eq.s (\ref{e-dgdtf}) and (\ref{e-dgdtc}) 
as follows:
\ber
\tau^{\rm diff}_{\rm flat} &=& \frac{4 \pi \sigma L^2}{c^2}, \\
\tau^{\rm diff}_{\rm curved} &=& \left(1 - \frac{2m}{r}\right)^{-1/2} \frac{4 \pi \sigma L^2}{c^2}, \\
\tau^{\rm adv}_{\rm flat} &=& \frac{L}{v(r)},\\
\tau^{\rm adv}_{\rm curved} &=& \frac{L}{v(r) + \frac{c^2}{4 \pi \sigma} \, \left(1 - \frac{2m}{r}\right)^{1/2} \, \frac{2m}{r^2}}. 
\eer
where $L$ is the length-scale associated with the underlying current distribution supporting the field. The extra factor in 
$\tau^{\rm diff}$ for the curved space-time is, of course, the well known red-shift factor introducing an overall slow 
down compared to the case of a flat space-time. In table (\ref{t-redshift}) we compare the values of the red-shift at 
the surfaces of the stars of different masses and for different equations of state. This factor is larger for a stiffer 
equation of state and for a larger mass - both of which makes the star more compact. But the numerical value of the 
red-shift factor is not very different from unity as can be seen from table (\ref{t-redshift}). It should be noted that
in defining $\tau^{\rm adv}$ we have simply collected terms proportional to the first space derivatives of $g(r,t)$.
Hence, the extra term in the numerator of $\tau^{\rm adv}$ for the curved space-time is actually a modification to the 
diffusive part due to the curvature of space-time and not really a modification to the convective part. But this definition
allows us to make upper bound estimates for the modifications to the final field values, due to the curvature of space-time,
as described below. \\

\begin{table}
\begin{center}{
\begin{tabular}{|l|l|r|r|} \hline
M/\msun & P71 & WFF & W74 \\ \cline{1-4}
1.4 & 1.499 & 1.263 & 1.23 \\ \cline{1-4}
1.6 & & 1.322 & 1.278 \\ \cline{1-4}
1.8 & & 1.396 & 1.335 \\ \cline{1-4}
2.0 & & & 1.408 \\ \cline{1-4}
\end{tabular}}
\end{center}
\caption[]{Red-shift factors at the surface of stars corresponding to the masses and the equations of state used 
in table (\ref{t-radius}).}
\label{t-redshift}
\end{table}

\ni It is evident from the above definitions that:
\beq
\tau^{\rm diff}_{\rm flat} < \tau^{\rm diff}_{\rm curved} \; \mbox{and} \;
\tau^{\rm adv}_{\rm flat} > \tau^{\rm adv}_{\rm curved}.
\eeq
This indicates that while the red-shift factor introduces an overall slow-down in the field evolution, the extra 
term, proportional to $\partial_r g(r,t)$ arising out of the curvature of space-time, accelerates the decay. Therefore,
the actual modification in the final surface field is given by the competition of these two effects. In order to make 
an estimate of this difference between the final surface field values obtained using a flat and a curved space-time, 
let us assume that $g(r,t)$ evolves in time with an approximate exponential behaviour with the above time-scales. In that 
case, the difference is obtained as:
\beq
\Delta = \frac{log \frac{B(t)_{\rm flat}}{B(0)_{\rm flat}} - log \frac{B(t)_{\rm curved}}{B(0)_{\rm curved}}}
{log \frac{B(t)_{\rm flat}}{B(0)_{\rm flat}}} = 1 - \frac{\tau_{\rm flat}}{\tau_{\rm curved}}\,,
\eeq
for either diffusion or convection. Before making estimates using the above definition we should note that the time
evolution of $g(r,t)$ does not really follow an exponential behaviour, particularly at late times. Therefore, 
the estimates that we make using these time-scales are nothing but very crude upper limits. \\

\ni For a star of mass 1.4~\msun (with an eos given by WFF, with the currents concentrated initially at a density of 
$10^{11}~\gcc$ and accreting at a rate of $10^{-10}$~\dmdt we find that 
$\Delta_{\rm diff} \sim 20\% \; \mbox{and} \; \Delta_{\rm adv} \sim 5\%$. This implies that, at most, we should see 
a difference of $\sim 15\%$. Consider curve 3 and 3a in fig.(\ref{fig02}) (which corresponds to these parameters).
We see a difference of about half an order of magnitude (before the curves level off and the effect of inner boundary
condition of `flux-freezing' starts controlling the evolution) which is close to $\sim 10\%$ of the corresponding value 
of the surface field and therefore within the upper limits estimated above. The actual difference between the final
surface fields are, of course, much smaller and is closer to $\sim 5\%$.

\section{conclusions}
\label{sconcl}

\ni From the results presented above we conclude that the effect of the curvature of space-time on the evolution 
of a crustal magnetic field of accreting neutron stars is quite small. To summarise:
\ben
\i the difference in the final field strength obtained by using a general relativistic calculation is much smaller
compared to the uncertainties already existing due to the plausible range of various crustal and binary evolution 
parameters;
\i the effect of curvature of space-time is larger for a stiffer equation of state or for a massive star but
such differences are again quite insignificant.
\een

\section*{Acknowledgments}
I would like to thank the organisers of ICGC-2000 (Kharagpur, India) for giving me the opportunity to present this 
work. Thanks are also due to T.~Padmanabhan, S.~Shankaranarayanan, Arun V.~Thampan for helpful discussions and to
the anonymous referee for raising several important points.

\beb
\bi Baym G., Pethick C., Pines D., 1969, Nat, 223, 673
\bi Baym G., Pethick C., Sutherland P., 1971, ApJ, 170, 299
\bi Bhattacharya D., 1995, {\em X-Ray Binaries}, ed. Lewin W.~H.~G., van Paradijs J., van den Heuvel E.~P.~J., 
    Cambridge University Press, pp. 233-251
\bi Bhattacharya D., 1996, {\em Pulsars : Problems and Progress}, 
    ed.  Johnston S., Walker M.~A., Bailes M., ASP Conference Series Vol.105, pp. 547-556
\bi Blandford R.~D., Applegate J.~H., Hernquist L., 1983, MNRAS, 204, 1025
\bi Brown E.~F., 2000, ApJ, 531, 988
\bi Choudhuri A.~R., Konar S., 2002, MNRAS, {\bf in press}, astro-ph/0108229
\bi Geppert U., Page D., Zannias T., 2000, PRD, 61, 123004
\bi Geppert U., Urpin V.~A., 1994, MNRAS, 271, 490
\bi Itoh N., Kohyama Y., Matsumoto N., Seki M., 1984, ApJ, 285, 758
\bi Jackson J.~D., 1975, {\em Classical Electrodynamics}, John Wiley \& Sons Inc., New York
\bi Konar S., 1997, Evolution of the Magnetic Field in Accreting Neutron Stars, {\em Ph.D. thesis}, 
    Indian Institute of Science, Bangalore
\bi Konar S., Bhattacharya D., 1997, MNRAS, 284, 311
\bi Konar S., Bhattacharya D., 1999a, MNRAS, 303, 588
\bi Konar S., Bhattacharya D., 1999b, MNRAS, 308, 795
\bi Konar S., Bhattacharya D., 2001,{\em The Neutron Star - Black Hole Connection}, 
    ed. Kouveliotou C., Ventura J., van den Heuvel E.~P.~J., Kluwer Academic Publishers, Dordrecht, pp. 71-76
\bi Konar S., Choudhuri A.~R., 2002, {\sl in preparation}
\bi Lorenz C.~P., Ravenhall D.~G., Pethick C.~J., 1993, PRL, 70, 379
\bi Negele J.~W., Vautherin D., 1973, Nucl.Phys., A207, 298
\bi Page D., Geppert U., Zannias T., 2000, A\&A, 360, 1052
\bi Pandharipande V.~R., 1971, Nucl.Phys.A, 178, 123
\bi Ruderman M.~A., 1972, ARA\&A, 10, 427
\bi Sengupta S., 1997, ApJ, 479, L133
\bi Sengupta S., 1998, ApJ, 501, 792
\bi Thompson C., Duncan R.~C., 1993, ApJ, 408, 194
\bi Urpin V.~A., Levshakov S.~A., Yakovlev D.~G., 1986, MNRAS, 219, 703
\bi van den Heuvel E.~P.~J., 1995, JA\&A, 16, 255
\bi Verbunt F., van den Heuvel E.~P.~J., 1995, {\em X-Ray Binaries}, ed. Lewin W.~H.~G., van Paradijs J., 
    van den Heuvel E.~P.~J., Cambridge University Press, pp. 457-494
\bi Ramachandra B.~S., Vishweshwara C.~V., 2000, Pramana, 55(4), 585
\bi Wald R.~M., 1984, {\em General Relativity}, University of Chicago Press
\bi Walecka J.~D., 1974, Ann.Phys., 83, 491
\bi Wasserman I., Shapiro S.~L., 1983, ApJ, 265, 1036
\bi Weinberg S., 1972, {\em Gravitation and Cosmology : Principles and Applications of the General Theory of Relativity},
    John Wiley \& Sons Inc., New York
\bi Wiringa R.~B., Fiks V., Fabrocini A., 1988, Phys.Rev., C38, 1010
\bi Yakovlev D.~G., Urpin V.~A., 1980, SvA, 24, 303
\bi Zdunik J.~L., Haensel P., Paczynski B., Miralda-Escude J., 1992, ApJ, 384, 129
\eeb


\end{document}